\documentclass[twocolumn,prl,superscriptaddress]{revtex4}
\usepackage{amsmath,amssymb,amsfonts,mathrsfs}
\usepackage{amsthm}
\usepackage{graphicx}
\usepackage{dcolumn}
\usepackage{bm}
\usepackage{color}

\begin{document}

\title{Emulating topological chiral magnetic effects in artificial Weyl semimetals}

\author{Xinsheng Tan}
\affiliation{National Laboratory of Solid State Microstructures, School of Physics,
Nanjing University, Nanjing 210093, China}
\author{Y. X. Zhao}
\affiliation{National Laboratory of Solid State Microstructures, School of Physics,
Nanjing University, Nanjing 210093, China}
\author{Qiang Liu}
\affiliation{National Laboratory of Solid State Microstructures, School of Physics,
Nanjing University, Nanjing 210093, China}
\author{Guangming Xue}
\affiliation{National Laboratory of Solid State Microstructures, School of Physics,
Nanjing University, Nanjing 210093, China}
\author{Hai-Feng Yu}
\email{hfyu@nju.edu.cn}
\affiliation{National Laboratory of Solid State Microstructures, School of Physics,
Nanjing University, Nanjing 210093, China}
\affiliation{Synergetic Innovation Center of Quantum Information and Quantum Physics,
University of Science and Technology of China, Hefei, Anhui 230026, China}
\author{Z. D. Wang}
\email{zwang@hku.hk}
\affiliation{Department of Physics and Center of Theoretical and Computational Physics,
	The University of Hong Kong, Pokfulam Road, Hong Kong, China}
\author{Yang Yu}
\email{yuyang@nju.edu.cn}
\affiliation{National Laboratory of Solid State Microstructures, School of Physics,
Nanjing University, Nanjing 210093, China}
\affiliation{Synergetic Innovation Center of Quantum Information and Quantum Physics,
University of Science and Technology of China, Hefei, Anhui 230026, China}

\begin{abstract}
We realized highly tunable Weyl semimetal-bands and subsequently emulated the topological chiral
magnetic effects in superconducting quantum circuits. Driving the superconducting
quantum circuits with elaborately designed microwave fields, we mapped the
momentum space of a lattice to the parameter space, realizing the
Hamiltonian of a Weyl semimetal. By measuring the energy spectrum, we
directly imaged the Weyl points of cubic lattices, whose topological winding
numbers were further determined from the Berry curvature measurement. In
particular, we used an additional microwave field to produce a
momentum-dependent chemical potential, from which the chiral magnetic
topological current was emulated in the presence of an artificial magnetic
field. This pure topological current is proportional to the magnetic field,
which is in contrast to the famous Ampere's law, and may have significant
impacts on topological materials and quantum devices.
\end{abstract}

\maketitle

\bigskip

Topological effects have attracted broad interests because they not only
deepen our understanding of fundamental physics but also are promising for
realizing robust quantum information processing. For instance, in
topological Weyl semimetals (WSM)~\cite{Xu-Science15,LvWeyl,Zhao-Wang-PRL15},
negative magneto-resistance effects induced by a so-called chiral anomaly
(CA) have recently been extensively addressed ~\cite%
{Son-Classical-NMR,NMR-Exper-1,NMR-Exper-2,NMR-Exper-3,NMR-Exper-4}.
However, the chiral magnetic effect (CME)~\cite{Fukushima_cme,KHARZEEV_cme},
which arises purely from the topology of paired Weyl points, has not yet
been detected. Here we have proposed and emulated CME in condensed
matter systems. We first realized an artificial Weyl semimetal band by using
superconducting quantum circuits \cite{you_QC,Clarke}. By driving the artificial system with
elaborately designed microwaves, we mapped the momentum space of a cubic
lattice to a controllable parameter space~\cite{nist,roushan,Gritsev,Tan-Maxwell,Tan_pt}.
By measuring the spectroscopy, we directly imaged the Weyl points, whose
winding numbers are further determined by the Berry curvature measurement.
Then we applied another microwave field for continuously tuning the chemical
potential to a well-designed momentum-dependent form, so that a unique
topological current originated from CME under an artificial magnetic field.
Moreover, we derived an elegant equation of the CME topological current in
our simulated Weyl semimetals with multiple Weyl pairs and manipulated the artificial currents by tuning various
parameters, ingeniously emulating CME.

A cubic lattice version of a simplified Weyl semimetal Hamiltonian may be
written as
\begin{equation}
H(k)=\sin k_{x}\sigma _{x}+\sin k_{y}\sigma _{y}+(\lambda +\cos k_{z})\sigma
_{z}+\mu _{eff}(\mathbf{k})\sigma _{0},  \label{ham2}
\end{equation}%
where $\sigma _{x,y,z}$ are Pauli matrices and $\sigma _{0}$ is the unit
matrix. \textbf{k} ($k_{x},$ $k_{y},$ $k_{z})$ is the momentum (or wave)
vector, and $\lambda $ ($|\lambda |\leq 1$) is an experimentally
controllable parameter that determines the $k_{z}$-coordinates of the Weyl
points (band crossing points). In most cases, chemical potential $\mu _{eff}(
$\textbf{k}$)$ is zero or a $k$-independent constant, which merely
introduces an extra global shift of the energy. A Weyl point can be
considered as a \textquotedblleft magnetic monopole\textquotedblright\ for
its associated Berry bundle. The monopole charge is defined as the Chern
number of the Berry bundle on a sphere enclosing the Weyl point in the
momentum space.
Physically, the monopole charge is a generalization of the chirality,
particularly a right (left)-handed Weyl point with a positive (negative)
unit charge. According to the Nielsen-Ninomiya no-go theorem~\cite{nn_nogo},
for a lattice model of Weyl semimetals, Weyl points always form dipole
constitutions of two oppositely charged monopoles. Thus the electromagnetic
response of WSMs lies essentially in that of a dipole. Namely, a left-handed
Weyl point is separated by $b_{\mu }$ ($\mu =0,1,2,3$) from a right-handed
one in the energy-momentum space. Due to the topological nature of the
dipole momentum $b_{\mu }$, a unique topological term responding exclusively
to the magnetic field may be expected from a minimal model of WSM consisting
of only one dipole of Weyl points. The corresponding action is given by
$S_{\Theta }=-\frac{1}{8\pi ^{2}}\int d^{4}x~\epsilon ^{\mu \nu \lambda
\sigma }b_{\mu }A_{\nu }\partial _{\lambda }A_{\sigma },$
where
$A_{\nu }$ represents the corresponding component of the electric-magnetic
potential, and $\epsilon ^{\mu \nu \lambda \sigma }$ is the anti-symmetric
tensor (here the electronic charge is set to $e=1$ for simplicity). It is
found that the above action leads to the anomalous current
$j^{\mu }=-\frac{1}{8\pi ^{2}}\epsilon ^{\mu \nu \lambda \sigma }b_{\nu
}F_{\lambda \sigma },
$ 
where $F_{\lambda \sigma }$ is the electromagnetic field tensor. A
remarkable result is that when two oppositely charged Weyl points are
separated by an energy difference $b_{0}$, a solely external magnetic field $%
\mathbf{B}$ (not a strong field for the sake of preserving band
characteristics) is able to induce an additional pure topological current $%
\mathbf{J}_{topo}$ given by
\cite{Notes-chern}
\begin{equation}
\mathbf{J}_{topo}=\frac{b_{0}}{4\pi ^{2}}\mathbf{B}.
\label{Chiral-magnetic-effect}
\end{equation}%
This is the equation of CME for one pair of Weyl points. The current arises
purely from a topological effect, and is directly proportional to the
magnetic field, in contrast to the famous Ampere's law.
Notably, this CME topological current
is inherently different from the CA
currents extensively addressed for WSMs over the past several years~\cite%
{Son-Classical-NMR,NMR-Exper-1,NMR-Exper-2,NMR-Exper-3,NMR-Exper-4}. Here,
the CA and CME currents are respectively originated from the two different
topological terms in the action
and described by distinct formulas~\cite{Notes-chern}.
The 
topological term 
in the action with the dipole momentum has the advantage of being readily
generalizable to a genetic WSM containing multiple dipoles. For a generic
case, we need to introduce a modified $b_{\mu }$
for multiple dipoles of monopoles: 
$b_{\mu }=\sum_{s}(K_{s}^{+,\mu }-K_{s}^{-,\mu })$ 
where $s$ labels left/right-handed Weyl points, and $K_{s}^{\pm ,\mu }$ are
the positions of left/right-handed Weyl points in the energy-momentum space.
It follows that CME can still be described by Eq.~(\ref%
{Chiral-magnetic-effect}) with the above-introduced modified $b_{0}$ 
\cite{Notes-chern}.


\begin{figure}[tbph]
\includegraphics[width=8.5cm]{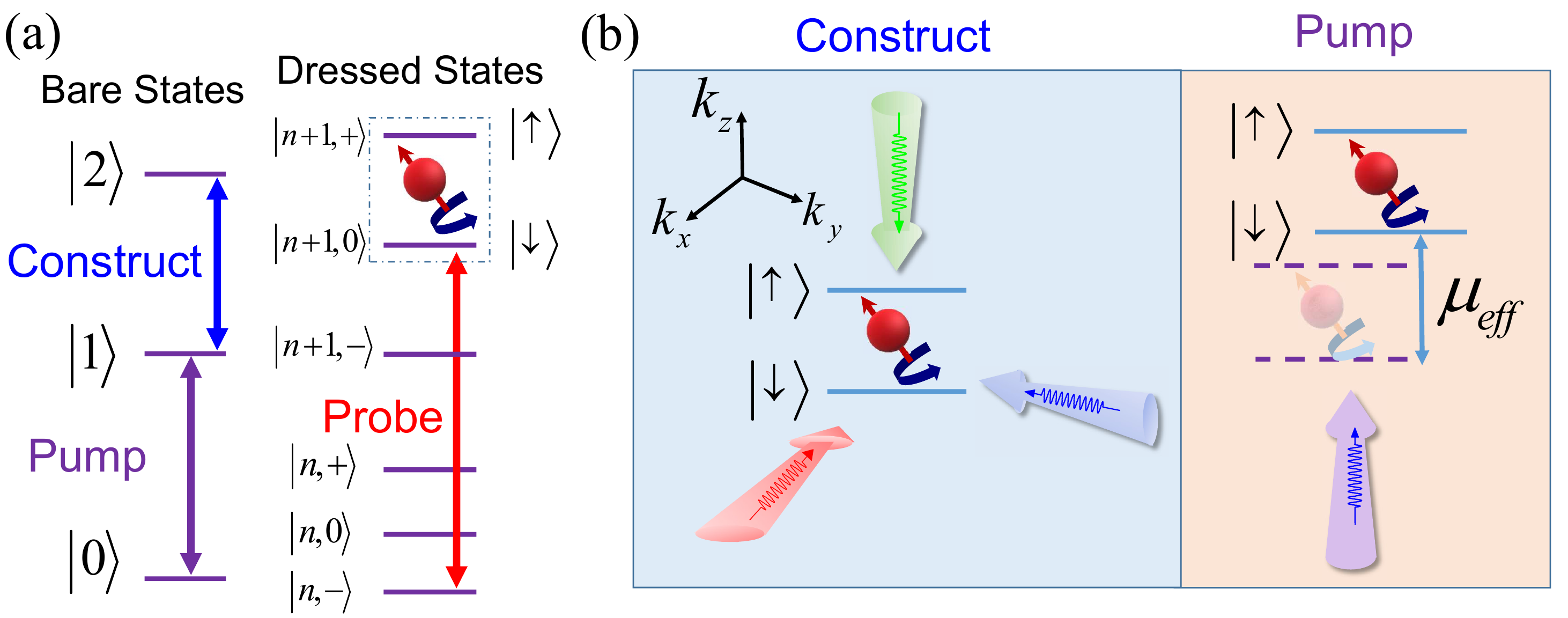}
\caption{Schematic of the energy level structure and Hamiltonian
manipulation for realizing WSM bands and CME in a superconducting transmon.
(a) The energy diagram of a transmon without (left) and with
(right) microwave driving. The lowest three energy levels of a transmon are
employed to construct the Hamiltonian in Eq. (\protect\ref{ham2}). The
construct (pump) microwave field is applied to $|1\rangle $ and $|2\rangle $
($|0\rangle $ and $|1\rangle )$. The system transforms to the dressed
states, as shown in the right panel. The top two states $|n+1,0\rangle $ and
$|n+1,+\rangle $ form a spin-1/2 system. The lowest state $|n,-\rangle $ is
selected as a reference level to detect spectrum. (b) Left panel:
microwaves with various frequencies, amplitudes, and phases are applied to
construct the Hamiltonian. In a rotating frame, the amplitudes and phases of
two construct microwaves contribute to $\protect\sigma _{x}$ and $\protect%
\sigma _{y}$, while the detuning defines $\protect\sigma _{z}.$ Right panel:
schematic for the effect of an extra pump microwave field, which is
to generate a parameter dependent energy offset. This is equivalent to
shifting the eigenenergy of the spin-1/2 particle, leading to a momentum
dependent chemical potential. }
\label{Demo}
\end{figure}

We used a three-dimensional (3D) superconducting transmon \cite%
{paik_3d,Tan-Maxwell} to realize WSM bands . The transmon, which
is composed of a single Josephson junction and two pads (250 $\mu $m $\times
$ 500 $\mu $m), was put in a rectangular aluminium cavity with the resonance
frequency of TE101 mode 9.026 GHz. In our experiments, the main function of
the cavity is to control and measure the transmon. The system is designed to
work in the dispersive regime. The whole sample package was cooled in a
dilution refrigerator to a base temperature of 30 mK. The dynamics of the
system can be described by the circuit QED theory which addresses the
interaction of an artificial atom with microwave fields \cite{Nori,blais_qed,wallraff,You_qed}. The
quantum states of the transmon can be controlled by microwaves. IQ mixers
combined with a 1 GHz arbitrary wave generator (AWG) were used to modulate
the amplitude, frequency, and phase of microwave pulses. The measurement was
performed with a \textquotedblleft high power readout" scheme \cite
{Reed_readout,Tan-Maxwell}. As a strong microwave on-resonance with the cavity is sent
in, the transmitted amplitude of the microwave reflects the state of the
transmon due to the non-linearity of the cavity QED system.

Conventionally, the transmon (coupled with cavity) exhibits anharmonic
multiple enengy levels. In our experiments, we considered the lowest three
bare energy levels of the transmon, as shown in the left part of Fig.~1(a),
namely, $|0\rangle $, $|1\rangle ,$ and $|2\rangle $. The transition
frequencies between different energy levels are $\omega _{10}/2\pi =$ 7.1755
GHz and $\omega _{21}/2\pi =$ 6.8310 GHz, respectively, which were
independently calibrated by saturation spectroscopy. To map the transmon
Hamiltonian to the form of Eq. (\ref{ham2}), we applied construct (pump)
microwaves to the transmon, coupling to states $|1\rangle $ and $|2\rangle $
($|0\rangle $ and $|1\rangle )$. The original bare states $\left\vert
0\right\rangle ,$ $|1\rangle ,$ and $|2\rangle $ will then transform to
microwave dressed states \cite{Tan_pt}. In the right part of Fig.~1(a), we show six relevant
dressed states, which are denoted by $\left\vert n,-\right\rangle $, $%
\left\vert n,0\right\rangle ,$ $\left\vert n,+\right\rangle ,$ $\left\vert
n+1,-\right\rangle $, $\left\vert n+1,0\right\rangle ,$ and $\left\vert
n+1,+\right\rangle ,$ respectively. Here $n$ is the average number of
photons in the coherent drive. The eigenenergies of these dressed states
depend on the microwave fields. In our experiments, the lowest level $%
\left\vert n,-\right\rangle $ acted as the reference level because the
system was always initialized in it. The top two levels $\left\vert
n+1,+\right\rangle $ and $\left\vert n+1,0\right\rangle $ were selected as
two eigen-states of spin-1/2 atoms. Under microwave driving, the effective
Hamiltonian of a spin-1/2 atom in the rotating frame may be written as $\hat{%
H}=\sum_{i=0}^{3}\Omega _{i}\sigma _{i}/2,\label{ham1}$ ($\hbar =1$ for
simplicity) where $\Omega _{1}$ $(\Omega _{2})$ corresponds to the frequency
of Rabi oscillations around the $x$ $(y)$ axis on the Bloch sphere, which
can be continuously adjusted by changing the amplitude and phase of the
construct microwave applied to the system. $\Omega _{3}=\omega _{21}-\omega
_{construct},$ is determined by the detuning between the energy level
spacing $\omega _{21}$ and the construct microwave frequency. $\Omega _{0}$,
corresponding to $\mu _{eff}$ in the above Hamiltonian, 
is related to the energy spacing between $\{\left\vert n+1,+\right\rangle
,\left\vert n+1,0\right\rangle \}$ and $\left\vert n,-\right\rangle $. Since
the splitting of $\left\vert n,-\right\rangle $ and $\left\vert
n,+\right\rangle $ ($\left\vert n+1,-\right\rangle $ and $\left\vert
n+1,+\right\rangle )$ depends on the frequency and amplitude of the pump
(construct) microwave, we can accurately design $\Omega _{0}$ after
calibration \cite{Notes-chern}. Therefore, by elaborately designing
the waveform of AWG and frequencies of microwaves, we can construct every
term in the above Hamiltonian 
points by points. We sent the probe microwave and swept the frequency. When
the frequency of the probe microwave matched the energy spacing between $%
\left\vert n,-\right\rangle $ and $\left\vert n+1,+\right\rangle $ (or $%
\left\vert n+1,0\right\rangle )$, a resonant peak could be observed. By
collecting the resonant peaks for various \textbf{k} parameters, we obtained
the band structures of Weyl semimetals.

\begin{figure}[tbph]
\includegraphics[width=8.5cm]{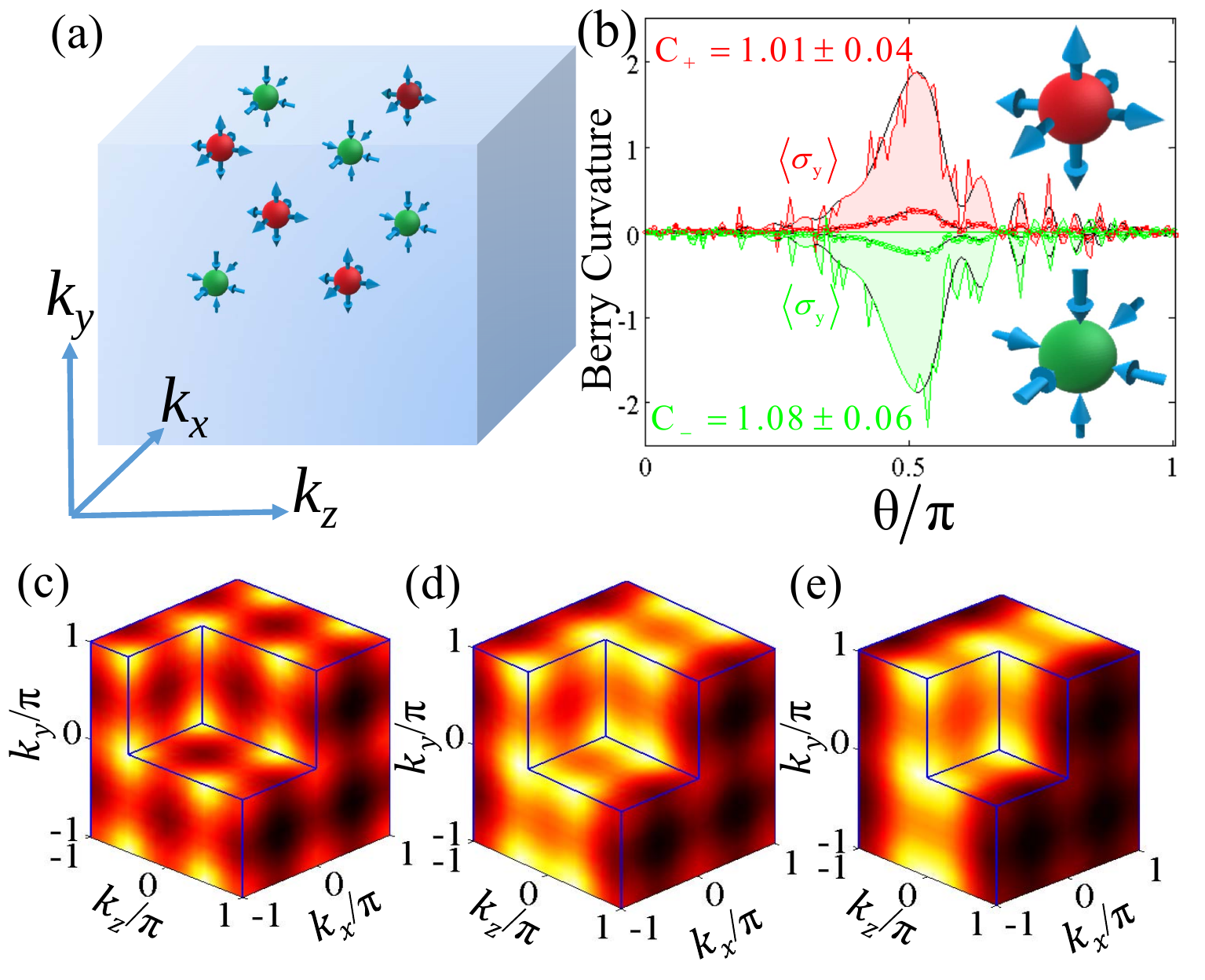}
\caption{ $\mathbf{a}$, Illustration of Weyl points in the first Brillouin
zone. Weyl points with positive (negative) charges are denoted by red
(green) color. (b) Schematic of the formation of winding numbers
at a positive (negative) Weyl point. Measured winding number from $\langle
\protect\sigma _{y}\rangle $ are close to +1 $(-$1), corresponding to Weyl
points with positive (negative) charges. (c), (d), (e)
Measured three-dimensional energy spectra of the Weyl semimetal in the
first Brillion zone for $\protect\lambda =0,$ $-0.5,$ $-1$. In order to show
the topological Weyl points, we cut the front corner of the spectrum. The
brightest points are Weyl points with the charges labeled. The locations of
gapless Weyl points are found to shift with $\protect\lambda $. When $%
\protect\lambda $ changes from 0 to $-$1, 8 Weyl points merge into 4 Dirac
points. }
\label{Spectrum}
\end{figure}

We first demonstrated the realization and manipulation of Weyl semimetal
bands for $\mu _{eff}($\textbf{k}$)=0$. In this case, the pump microwave
field was turned off. The corresponding waveforms were synthesized by AWG
and sent to the microwave sources for constructing the Hamiltonian in Eq.~(1). Here we set $\Omega _{1}/2=\Omega \sin k_{x}$, $\Omega _{2}/2=\Omega
\sin k_{y}$, and $\Omega _{3}/2=\Omega (\lambda +\cos k_{z})$, with unit $%
\Omega $ = 10 MHz. To visualize the band structure of the Weyl semimetal, we
measured the spectroscopy of the lattice Hamiltonian in the first Brillouin
zone, as shown in Fig.~\ref{Spectrum}. Weyl points, as the signature of Weyl
semimetals, are directly observed. To the best of our knowledge, this is the
first experiment that directly images Weyl points in a momentum spectrum of cubic lattices.
As illustrated in Fig.~\ref{Spectrum}, eight Weyl points are observed in the
first Brillouin zone. There are four points with a positive (negative)
charge, which are named $W_{+}$ $(W_{-})$ and denoted by red (green) color.
To characterize their topological properties, we also measured the winding
number of each Weyl point by detecting the Berry curvature, which was
determined by measuring the non-adiabatic response to the change of the
external parameter \cite{Gritsev,roushan,nist}. We let the system evolve
quasi-adiabatically along a designed path in the parameter space, so that
the Berry curvature is directly related to $\langle \sigma
_{y}\rangle $ \cite{Notes-chern}. Then the winding numbers for $%
W_{\pm }$ could be obtained by performing quantum state tomography. As shown
in Fig.~\ref{Spectrum}(b), the winding number of $W_{\pm}$
approximates $\pm 1$, close to the predicted value, where $\pm $ of the
winding number corresponds to the topological charge sign of Weyl points.
Moreover, we can continuously vary $\lambda $ to explore various topological
properties of Weyl semimetals. For instance, we observed that with the
change of $\lambda $ Weyl points merge and annihilate. Shown in Fig. \ref%
{Spectrum} (c)-(e) are three dimensional spectra for $\lambda $ = 0, $-$0.5,
and $-$1, from left to right. When $\lambda $ = 0, there are eight Weyl
points, locating at $(0,$ $0,$ $\pm \pi /2),$ $(0,$ $\pi ,$ $\pm \pi /2),$ $%
(\pi ,$ $0,$ $\pm \pi /2),$ and $(\pi ,$ $\pi ,$ $\pm \pi /2)$,
respectively. With $\lambda $ decreasing from 0 to $-1$, Weyl points move
along the $k_{z}$ direction. When $\lambda =-1$, the eight Weyl points merge
to four Dirac points, which corresponds to a topological phase transition
from the topological WSM phase to a normal Dirac metal phase.

Having realized a tunable WSM band, we can turn on the pump microwave field
to generate topological currents in the system, which leads to the
realization of CME. From Eq.~(\ref{Chiral-magnetic-effect}), we noted that
there are two facters that account for CME topological currents. The first
is a non-zero $b_{0}$, which is difficult to fulfill experimentally, and the
second is the presence of a magnetic field. We firstly show how to obtain a
finite $b_{0},$ which is essential in our simulation. As shown in Eq.~(1), $%
\mu _{eff}($\textbf{k}$)$ is the prefactor of the $\sigma _{0}$ term, which
corresponds to the chemical potential of the system. For our WSM, we have
four pairs of Weyl points. In an ideal case, $b_{0}$ may be four times the
chemical potential (or energy) difference between a single pair of $W_{+}$
and $W_{-}$ points. We noticed that the Weyl points have different locations
in momentum space. If one can produce a momentum dependent chemical
potential, the energy difference of the two Weyl points will be
non-vanishing. In order to realize a momentum dependent $\mu _{eff}($\textbf{%
k}$)$, we applied an extra pump microwave field to shift the reference level
for different \textbf{k}. In the dressed state picture, the absolute value
of the eigenenergy is determined by the energy spacing between subspace $%
\{\left\vert n+1,+\right\rangle ,\left\vert n+1,0\right\rangle \}$ and $%
\left\vert n,-\right\rangle $, which is $\omega _{01}+\Delta _{01}/4+3\sqrt{%
\Delta _{01}^{2}+\Omega _{01}^{2}}/4$~\cite{Notes-chern}.
Here $\Delta _{01}$ is detuning between level spacing $\omega _{10}$ and the
pump microwave, while $\Omega _{01}$ is the coupling Rabi frequency, which
is proportional to the amplitude of the pump microwave. In contrast to the
original value $\omega _{01}$, the magnitude of energy shift is $\Delta
_{01}/4+3\sqrt{\Delta _{01}^{2}+\Omega _{01}^{2}}/4$, which can be
accurately controlled by adjusting the amplitude and frequency of pump
microwaves. For instance, using an accurately designed waveform, we could
obtain a special kind of
\begin{equation}
\mu _{eff}(\mathbf{k})=f(k_{z})\cos (k_{x})\cos (k_{y}),  \label{mu}
\end{equation}%
where $f(k_{z})$ is a $2\pi $-periodic function of $k_{z}$, which can be set
as $f(k_{z})=\alpha k_{z}/(10\pi )$ for $|k_{z}|\leq 7\pi /8$, while $%
f(k_{z})=-7\alpha /10\pi \lbrack k_{z}-sign(k_{z})]$ for $7\pi /8\leq
|k_{z}|\leq \pi $ 
in the range of $[-\pi ,\pi ]$, with $\alpha $ being a preset parameter. In our experiments, $f(k_{z})$ in the range $[-7\pi /8,7\pi /8]$
is a simple linear function of $k_{z},$ which can generate the energy
difference $\mu _{eff}$ of a relevant pair of Weyl points. The function in $%
[-\pi ,-7\pi /8)$ and ($7\pi /8,\pi ]$ is chosen to have the periodical
boundary condition
. In Fig.~3(b) we show an example of $\mu _{eff}$ with $k_{x}=k_{y}=0$, $\alpha =1
$. Notably, our specific choice of the $\cos (k_{x})\cos (k_{y})$ term in $%
\mu _{eff}$ is to ensure that the total current contributed from four pairs
of Weyl points is four times that for a single pair.

\begin{figure}[tbph]
\includegraphics[width=8.5cm]{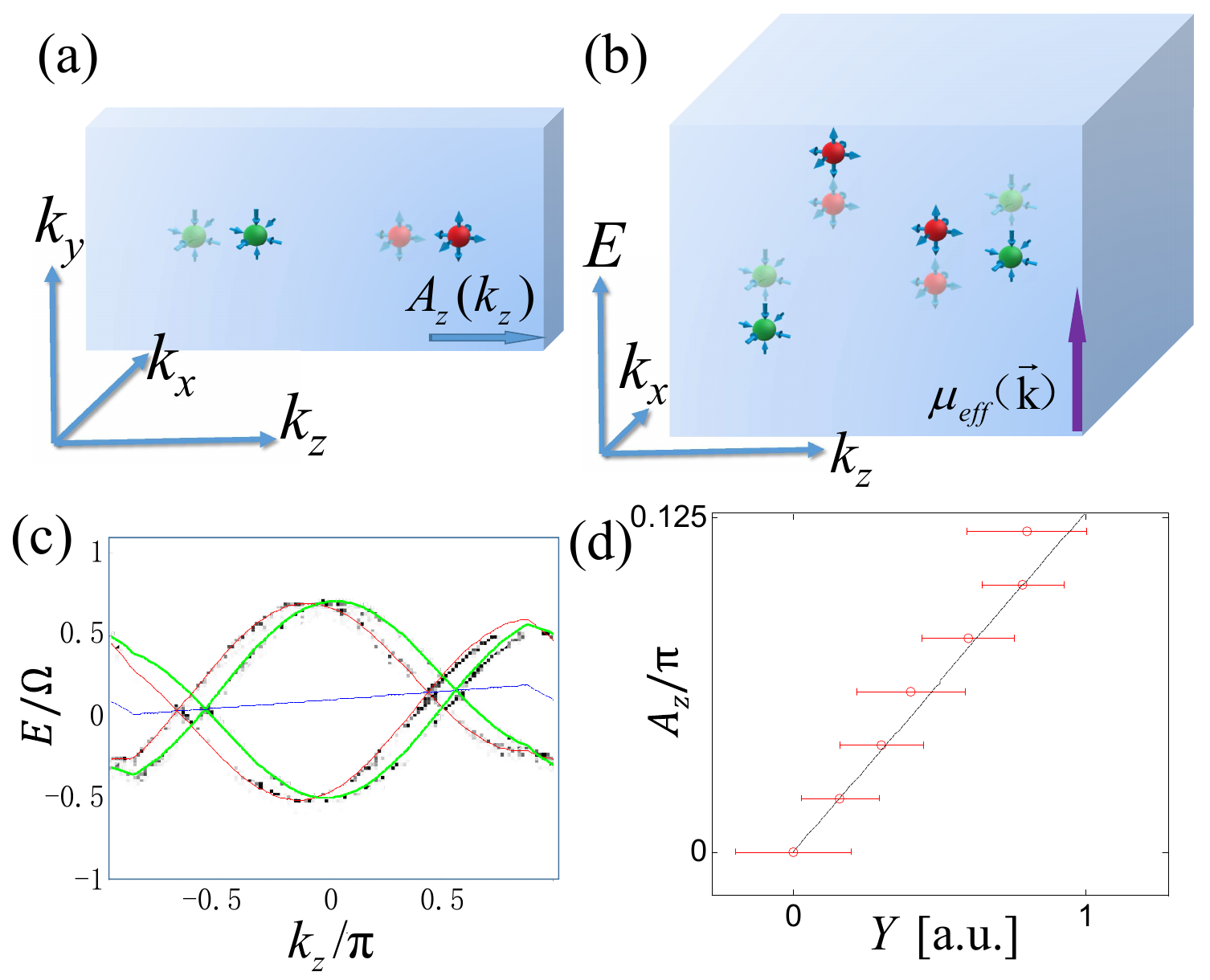}
\caption{\ Effects of an offset $A_{z}$ and momentum dependent $\protect\mu %
_{eff}$ are illustrated in (a) and (b). (c)
Spectrum along the $k_{z}$ direction for Hamiltonian in Eq.~(1) for $\protect%
\mu _{eff}(\mathbf{k})$ = $f(k_{z})\cos (k_{x})\cos (k_{y})$ (blue line)
with $k_{x}=k_{y}=0$ , where $f(k_{z})$ is given in Eq.~(\protect\ref{mu})
with $\protect\alpha =1$. The vector potential $A_{z},$ which causes the
shift of the spectrum, is 0 (green) and $\protect\pi /8$ (red) respectively.
By adjusting $A_{z}$, we can obtain different $\Delta Y$ (from $\Delta E)$.
With the maximal $A_{z}$ ($\protect\pi /8$ here), $Y$ is set as 1. (d)
Plot of $A_{z}$ as a function of $Y$. Since $B_{x}\propto A_{z}/Y$, we
extract the artificial magnetic field $B_{x}$ from the slope of the best
fit. Error bars are derived from the width of resonant peaks in the
spectrum. }
\label{Potential}
\end{figure}

We now turn to the second factor of CME. From Eq.~(\ref%
{Chiral-magnetic-effect}), the CME topological current will be proportional
to both $b_{0}$ and 
the magnetic field $\mathbf{B}$. Therefore, we can directly determine a
topological current by measuring $b_{0}$ from the spectrum data for any
finite $\mathbf{B}$. Utilizing the high controllability of the
superconducting circuits, we may produce an artificial magnetic field.
Since $\mathbf{B}=\nabla \times $ \textbf{A}, we can introduce an artificial
vector potential \textbf{A} to generate the artificial magnetic field
needed. Without a loss of generality, we assumed that \textbf{B} was along
the $x$ direction and \textbf{A} had only the $z$-component, such that $%
B_{x}=\partial A_{z}/\partial Y$, where a set of $A_{z}$ in our experiment
was chosen to be proportional to a fictitious coordinate $Y$ (with the
dimensionality of length). Considering the form of the canonical momentum,
we can introduce the vector potential by adding a controllable shift on the
momentum $k_{z}$, namely, we transformed $k_{z}$ to $k_{z}+A_{z}$ in the $%
\sigma _{z}$ prefactor of Eq.~(\ref{ham2}), resulting in $\Omega
_{3}/2=\Omega (\lambda +cos(k_{z}+A_{z}))$. In the presence of this offset,
the artificial magnetic field may be introduced to act on $e$-charge single
particles, and thus the CME topological current will also be generated.
In order to extract $B_{x}$, we here adopted a scenario of fictitious work
to evaluate $\partial A_{z}/\partial Y$ from our experimental data. Since we
had a good linear momentum dependent shift of the energy bands near the Weyl
points, a small change of $A_{z}$ will cause an energy difference in $\Delta
E$, i.e., $\Delta E\propto \Delta A_{z}$ near the Weyl points. On the other
hand, we may write $\Delta E=F_{Y}\Delta Y$ near one Weyl point, where $%
F_{Y} $ could be viewed as a constant fictitious force acting on the
particle which moved $\Delta Y$ along the $Y$-direction. Therefore, $\Delta
A_{z}\propto \Delta Y$, with the prefactor defining the magnitude of $B_{x}$%
. In our experiments, by choosing various small $A_{z}$ near the Weyl point,
we measured the corresponding $\Delta A_{z}$ and $\Delta E$ (i.e., $\Delta Y$
if we set $F_{Y}=1$ force unit). By plotting $A_{z}$ as a function of $Y$,
we indeed observed a good linear relation, whose slope may approximately
give the artificial field $B_{x}$. Since a key objective of our simulation
is to demonstrate CME, obtaining the accurate absolute value of $B_{x}$ is
not essential. For various fixed $B_{x}$, we measured $b_{0}$, from which we
were able to determine the CME topological current $\mathbf{J}_{topo}$ based
on our derived Eq.~(\ref{Chiral-magnetic-effect}).

\begin{figure}[tbph]
\includegraphics[width=8.5cm]{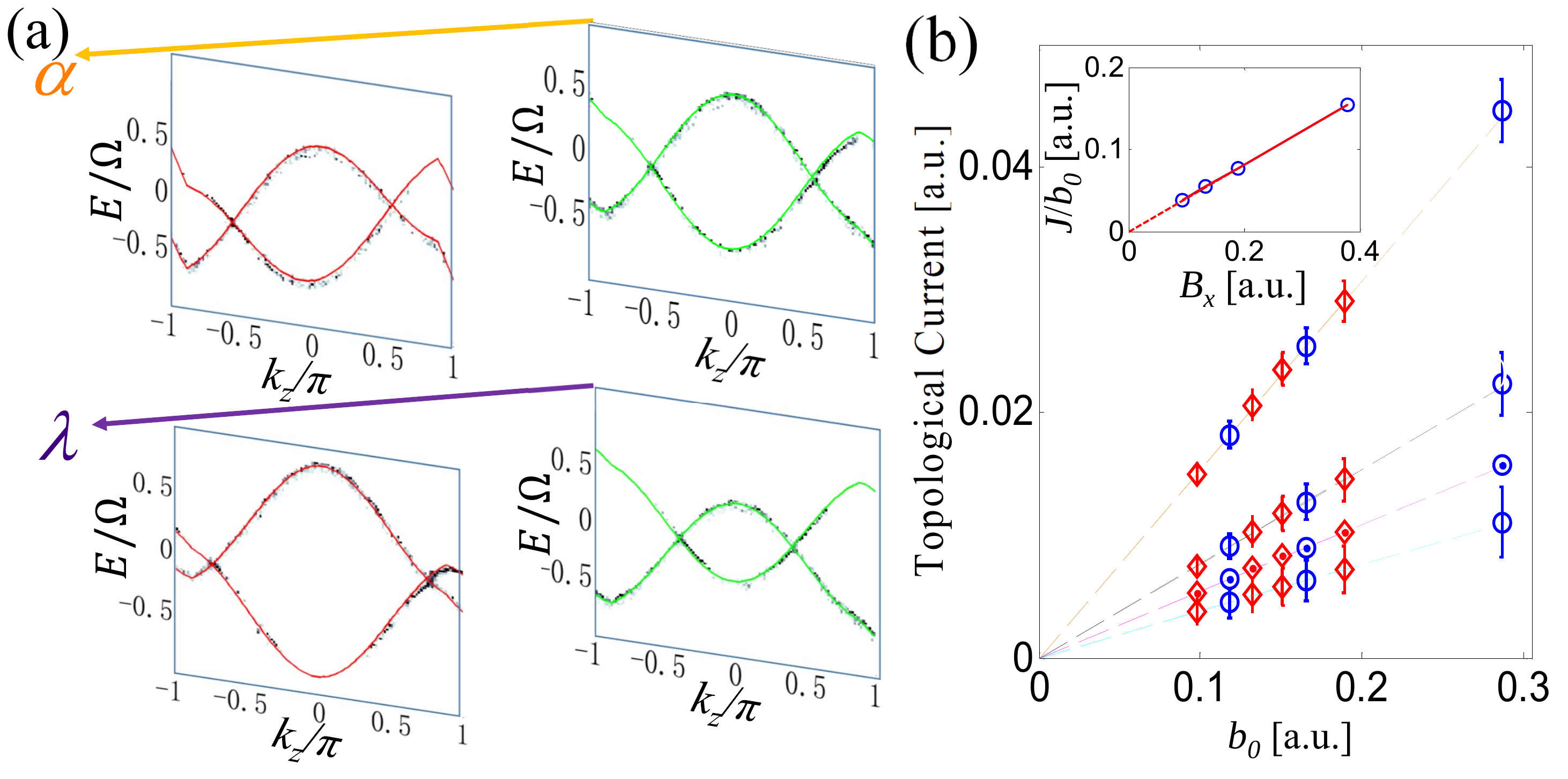}
\caption{Probe of CME topological currents using two different methods.
(a) Energy spectrum with various $b_{0}$ obtained by
changing $\protect\lambda $ and $\protect\mu _{eff}$. The upper panel shows
the spectrum for different $\protect\alpha $ of $\protect\mu _{eff}$ with $%
\protect\lambda =0$ : $\protect\alpha =1$ (green) and $\protect\alpha =3$
(red). The lower panel shows the spectrum for $\protect\lambda =$ 0.5 (red)
and $-0.5$ (green) with $\protect\alpha =1 $ . Solid dots and lines are
experimental data and theoretical calculations respectively. (b)
The total topological current, which is contributed to by all four pairs of
Weyl points, is plotted as a function of $b_{0}$ for various $B_{x}.$ From
bottom to top, $B_{x}$ is 0.09, 0.13, 0.19, and 0.38, respectively. Blue and
red circles are experimental data obtained from the shifts of $\protect%
\lambda $ and $\protect\alpha $, respectively, while dashed lines are
theoretical results. The inset of (b) shows the normalized
topological current $J/b_{0}$ as a function of $B_{x}$ for the two methods. }
\label{topo_current}
\end{figure}

One advantage for simulating topological band structures by using
superconducting transmon is the full controllability of the parameters
space. We can manipulate the Hamiltonian in Eq.~(1) by continuously tuning
parameters $\lambda ,$ $A_{z}$, and $\mu _{eff}(\mathbf{k}).$ For instance,
there are two ways we can vary the CME topological current. One is to change
$\lambda $ to increase the distance between the two Weyl points with the
slope of $\mu _{eff}(\mathbf{k})$ fixed. The other is to tune the slope of $%
\mu _{eff}(\mathbf{k})$ while keeping a constant $\lambda $. From Eq.~(3),
we know that the slope can be tuned by changing $\alpha $. As indicated in
Fig.~\ref{topo_current}(a), with $\lambda $ varying from 0.5 to $-$0.5 for $%
\alpha =1$, the positions of Weyl points shift, causing the energy
difference to change from 0.0813 GHz (green) to 0.1512 GHz (red). On the
other hand, when $k_{z}$ is in the range of (-$7\pi /8$, $7\pi /8$), to
changing the slope of $f(k_{z})$ (i.e., $\alpha $) in $\mu _{eff}(\mathbf{k})
$ also modifies the energy difference of two Weyl points for a fixed $%
\lambda =0$ : the energy difference is 0.151 GHz for $\alpha =1$ (green),
while it is 0.430 GHz for $\alpha =3$ (red). Both methods effectively
changed $b_{0}$, and hence the topological current. This current varied with
the magnetic field, indicating that we have successfully introduced an
artificial magnetic field and generated the topological current. For various
$B_{x},$ we plotted the topological currents as a function of $b_{0}$ for
both methods. The currents extracted from the two approaches fall on one
straight line for the same $B_{x}$, as shown in the main panel of Fig.~4(b).
Moreover, the normalized topological current $J/b_{0}$ for two different
methods depends solely on $B_{x}$, as seen from the inset of Fig.~4(b).

Our work presents the first experimental report on quantum
emulation of CME, paving the way for further exploration and simulation of
topological physics. Our work also modified the conventional Ampere's law
with regard to WSM-like real materials with multiple Weyl pairs, which may have an important impact on
fundamental physics and material science.

\begin{acknowledgments}
\textit{Acknowledgments} This work was partly supported by the the NKRDP of
China (Grant No. 2016YFA0301802), NSFC (Grant No. 91321310, No. 11274156,
No. 11504165, No. 11474152, No. 61521001), the GRF of Hong Kong (Grants No.
HKU173055/15P and No. HKU173309/16P).
\end{acknowledgments}

\end{document}